\documentclass[12pt,preprint]{aastex}

\usepackage{epsfig}
\usepackage{graphicx,amsmath}
\usepackage{graphicx}
\usepackage{amsmath}
\usepackage{amssymb}
\usepackage{color}
\shorttitle{A RECIPE TO PROBE ALTERNATIVE THEORIES OF GRAVITATION}
\shortauthors{Brandao \& de Araujo}

\begin{document}
\title{A RECIPE TO PROBE ALTERNATIVE THEORIES OF GRAVITATION VIA N-BODY NUMERICAL SIMULATIONS. I. SPIRAL GALAXIES}

\author{C. S. S. BRANDAO
and J. C. N. DE ARAUJO}
\affil{Divis\~{a}o de Astrof\'{i}sica, Instituto Nacional de Pesquisas Espaciais, \\
S. J. Campos, SP 12227-010, Brazil;}
\email{claudiosoriano.uesc@gmail.com, jcarlos.dearaujo@inpe.br}
\begin{abstract} 
A way to probe alternative theories of gravitation is to study if they could account for the structures of the universe. We then modified the well-known Gadget-2 code to probe alternative theories of gravitation through galactic dynamics. As an application, we simulate the evolution of spiral galaxies to probe alternative theories of gravitation whose weak field limits have a Yukawa-like gravitational potential. These simulations show that galactic dynamics can be used to constrain the parameters associated with alternative theories of gravitation. It is worth stressing that the recipe given in the present study can be applied to any other alternative theory of gravitation in which the superposition principle is valid.
\end{abstract}


\section{Introduction} 
\label{intro}

One of the most important challenges of modern cosmology concerns the dark energy problem.
The nature or origin of dark energy cannot be associated with particles, and its interpretation based on quantum field theories does not provide a satisfactory answer \citep{cpt1992}. The scalar field hypotheses, as given by quintessence models, do not either solve the problem satisfactorily, but instead, give rise to other questions without explanations. Moreover, observations claim that $\sim 70 \%$ of the total energy composition of the universe is made of this puzzling (cosmological) ingredient that accelerates the expansion of the universe \citep{perl}, like an antigravity term in the Einstein's equations.

Another puzzling ingredient of the Cosmos is the dark matter \citep{steven}. Cosmology shows us, based on observations of galactic and cluster dynamics, gravitational lenses, etc., that the dark matter composition of the universe is $\sim 25\%$. In principle, the dark matter nature can be explained by particle physics; one expects that the large hadron collider helps us to give a good answer to this issue.

Recently, there appear in the literature some new alternative theories to the Einstein's General Relativity aiming to explain the structures
of the universe without dark contents (matter and energy). They are based on different hypotheses (e.g., massive gravitons, scalar-tensor theories of gravity, etc.) and they do not have in the weak field limit the Newtonian gravitational potential
\cite[see, e.g.,][]{moffat96,piazza03,rodriguez-meza05,sign05,araujo2007}.

In some theories, for example, the gravitational potential is Yukawa-like (hereafter Yukawian gravitational potential, YGP) in the weak field limit. The YGP reads

\begin{equation}
\label{ygp}
\phi=-\frac{Gm}{r}e^{-r/\lambda},
\end{equation}
where $m$ is the point-mass source of the field, $r$ is the distance to the point mass, and $\lambda$ is a characteristic length. In some theories, one postulates the existence of a massive boson called graviton, of mass $m_{\rm g}$, whose Compton wavelength can be interpreted as the $\lambda$ parameter.

It is worth noting that many authors consider that to probe a given alternative potential
it is enough to reproduce the rotation curves of spiral galaxies assuming centrifugal equilibrium.
In fact, the best way to probe a potential is to use a galactic dynamical approach. Moreover,
it is necessary to verify, even using simulations (living systems), not only the rotation
curves but also the surface density profiles. From the observations one can infer that this profile is
exponential, therefore a given alternative potential must be consistent with it too.

Note also that alternative gravitational potentials are basically of two types: those in which
the superposition principle applies and those in which it does not.
For the first type, one can use $N$-body simulations, in which the use of the superposition
principle is inherent. For the second type one must use, for example, smoothed particle hydrodynamics simulations,
where it is not necessary to be concerned with the superposition principle. Modified Newtonian dynamics, for instance, is of the second type, therefore one cannot use $N$-body simulations to
model mondian galaxies.

In our previous papers \citep{brandaoearaujo2009a,brandaoearaujo2009b}, we modified and tested an $N$-body code replacing the Newtonian potential by the YGP. Moreover, we made some numerical simulations to study elliptical galaxies.
It is worth mentioning that although we adopted this particular potential, the recipe given in these papers can be applied to any alternative gravitational potential in which the superposition principle is valid.

To complete our studies of galactic systems with alternative gravitational potentials, we consider in the present paper the modeling of late-type systems. Therefore, our main aim here is to give a recipe to perform such a study, and we adopted again the YGP as an example.

Concerning in particular the YGP, in almost all previous works concerning it, the investigations had been made using analytical or numerical approaches. For example,  \citet{araujo2007} have probed how the YGP can change the rotation curves of spiral galaxies, using ``static'' (centrifugal equilibrium) and analytic models of exponential disks.

In the present work, as an alternative to the approach used by de Araujo \& Miranda, we study numerical $N$-body simulations of disk galaxies and investigate how the YGP changes the canonical morphology of a simulated disk.

We show in the end how ``living'' systems behave under the YGP. Verifying therefore whether the YGP can or cannot keep the disk galaxies in dynamical equilibrium.

The paper is organized as follows: in Section 2, we present the galactic model used and the numerical code adopted to perform the simulations; in Section 3, we present the simulations and discuss the results; and, finally, in Section 4 we present the main conclusions.

\section{METHOD AND SCENARIO: A SPIRAL GALAXY MODEL UNDER YGP} 
\label{metodo}

To investigate YGP at galactic scale, or any other alternative potential in which the superposition principle can be applied, one has to choose a typical model (dynamical equilibrium) for the galaxies, and write an efficient $N$-body code, based on the tree method \citep{barnesehut} for example, to follow the evolution of the galaxies.

In the present paper we will consider disk galaxies. We know that such galaxies, their observational, structural and dynamical properties are well understood by the galactic dynamics approach \cite[see, e.g.,][]{bm1998,bt2008}. These systems are well modeled by numerical simulation tools and their secular evolution, under the mutual forces among their particles, can be followed \cite[see, e.g.,][and references therein]  {hernquist1993,springel1999,sdmh2005,atana1,atana2}.

In the present paper, our model of disk galaxies is based on the models described by \citet{springel1999} and on some other implementations introduced by \citet{sdmh2005}. Our galaxy model is called, throughout this paper, Springel-Di Matteo-Hernquist disks (hereafter SdMH disks), due to their numerical prescriptions, developed specially to model late-type systems. However, we here simplify the SdMH disks, since we do not include bulges or gas particles, due to the fact that our investigation is aimed only to study the overall dynamical and gravitational properties of disk galaxies under alternative potentials.

It is worth noting again that, although we consider here the particular case of the YGP, the prescription considered throughout the present paper holds for any other alternative potential one would like to probe, since they respect the superposition principle.

\subsection{Modeling Galaxies with an $N$-dody Code} 
\label{code}

To simulate a spiral galaxy under YGP, we have chosen the Gadget-2 code \citep{springel2005} and changed its structure to include the YGP, as we show in detail in our previous work \citep{brandaoearaujo2009a}.

Gadget-2 is based on the tree code method \citep{barnesehut}. So, its computational effort is $ \emph{O}(N \log(N))$, instead of $\emph{O} (N^2)$ operations required by direct sum algorithms. With this kind of code \cite[see][for details]{brandaoearaujo2009a}, we can integrate all equations of motion of a set of $N$ collisionless particles and follow their evolution, as we did for example for elliptical galaxies \citep{brandaoearaujo2009b}.

We recall here our previous arguments to justify our methodology. We
constructed galaxies initially with a Newtonian potential and then we submit them to the YGP. It is important to emphasize that particles represent physical observable quantities, such as positions, velocities and masses distributed over a given volume. So, when we realize an initial galaxy snapshot with the Newtonian potential, we mimic the following observational characteristics: radial luminosity profile, radial density profile, and velocity dispersions. It is important to bear in mind that Newtonian galaxies are consistent with observations, if dark matter is taken into account
\cite[see, e.g.,][]{oh2011,guedes2011}.

With this philosophy, independently on the physics used to build up galaxies, the simulated particles must reproduce the observed characteristics of real objects. Our aim is then to check, at the end of our simulations for a given alternative gravitational potential, if these characteristics are really consistent with observed objects.

One could argue that the best way to model Yukawian galaxies would be to consider their formations starting with collapsing halos. But since, as mentioned above, we start with a Newtonian model, this procedure could in principle be considered a limitation of the present approach. In the near future, however, we intend to investigate alternative potentials starting with collapsing halos.

\subsection{Assembling Galactic Halos and Disks: The particle's Positions} 

Our late-type system is composed by a dark matter halo modeled by a Hernquist sphere and an exponential-\textit{Spitzer} disk. The halo density distribution law reads
\begin{equation}
\rho_{\rm{dm}}(r) = \frac{M_{\rm{dm}}}{2\pi} \frac{a}{r(r+a)^3}\, ,
\label{hernquistdensity}
\end{equation}
where $\rho_{\rm{dm}}(r)$ is the radial density profile of the dark matter ($\rm{dm}$) halo, $M_{\rm{dm}}$ is the halo mass,  $r$ is the radial distance from the center of the whole mass distribution, and $a$ is a characteristic length of the halo's core. This parameter is related to the concentration parameter $c$ of the Navarro, Frenk and White (hereafter NFW) sphere \citep{nfw1,nfw2} of concentration index $c = r_{200}/r_s$, where $r_{200}$ is the \textit{virial radius}, where the mean overdensity, as compared to the critical density, is 200. The $r_s$ parameter is a characteristic scale length of the NFW sphere.

Recall that the scale lengths of the Hernquist sphere and the NFW sphere are related as
\begin{equation}
a = r_s \sqrt{2[\ln (1+c)-c/(1+c)]}.
\label{aandrs}
\end{equation}

Also, recall that the mass of the NFW sphere diverges as $r\rightarrow \infty$, while the mass of the Hernquist spheres converges as $r\rightarrow \infty$ to the value $M_{\rm{dm}}$. The two profiles agree very well within the virial radius $r_{200}$. But NFW spheres must be truncated, while Hernquist spheres do not. In this way, we follow \citet{sdmh2005} and use the Hernquist sphere to model the dark matter halo.

In the present work, we consider a disk made of baryonic particles. Our model does not consider gas, star formation, wave shocks, supernovae or supermassive black holes. We have also considered that the baryonic mass is a fraction of the total mass  $M_{\rm{disk}} = m_d M_{\rm{tot}}$, where $m_d$ is a dimensionless parameter, $M_{\rm{disk}}$ is the disk mass, and the total mass is $M_{\rm{tot}} = M_{\rm{disk}} + M_{\rm{dm}}$.

The disk has a residual spin from its primordial halo, from which it was formed. The spin parameter $\lambda_s$  reads

\begin{equation}
\lambda_s = \frac{J |E|^{1/2}}{GM^{5/2}},
\label{spinparameter}
\end{equation}

\noindent where $J$ is the total angular momentum of the primordial halo, $E$ is the total energy, $G$ is the universal gravitational constant, and $M=M_{\rm{dm}} + M_{\rm{disk}}$ is the primordial mass. The spin parameter is used to compute the parameter $f_c$ that will be defined below.

The disk density distribution is given by an exponential law and the \textit{Spitzer} isothermal sheet, namely:
\begin{equation}
\rho_{d}(R,z) = \frac{M_{d}}{4 \pi z_0 h^2}\exp\left(- \frac{R}{h}\right) {\rm sech}^2 \left( \frac{z}{z_0} \right),
\label{diskdensity}
\end{equation}
where $z_0$ is the disk thickness, $h$ is the radial scale length, and $M_{d}$ is the total disk mass. Differently from \citet{sdmh2005}, that left $z_0$ as a free parameter, we followed \citet{springel1999} and set $z_0 \simeq 0.2 R_d $
due to the fact that many late-type systems have this typical scale length. All particle's positions are distributed by the Monte Carlo method.

\subsection{Assembling Galactic Halos and Hisks. The Particle's Velocities} 
\label{assembly}

The main ingredient of this subsection has to do with how to distribute velocities of the halo and disk particles. While the entire prescription is found in the literature, we recall here only the most important ingredients to model our galaxies.

The velocity structure of our model depends on the calculation of the potential generated by the matter. Hernquist spheres have a potential given by
\begin{equation}
\Phi_{\rm{dm}} = - \frac{G M_{\rm{dm}}}{r + a}.
\label{hernquistpotential}
\end{equation}

To calculate the potential from the disk $\Phi_d$, contrary to \citet{springel1999} and \citet{sdmh2005}, we follow Equation (2.170) from \citet{bt2008}, who consider an exponential thick disk of a completely flattened homoeoid.

The calculation of the potentials is a key issue, since it is used to compute the velocity dispersions. For this axisymmetric system, we assume that the velocity distribution function is $f(E,L_z)$ \cite[see, e.g.,][]{magorrian}, where $E$ is the total energy and $L_z$ is the $z$-component of the angular momentum.

It follows that the first velocity moments are given by (in cylindrical coordinates)

\begin{equation}
\overline{v_R} = \overline{v_z} = \overline{v_R v_z} = \overline{v_z v_\phi} = \overline{v_R v_\phi} = 0,
\label{dispsnulos}
\end{equation}

\begin{equation}
\overline{v^2_R} = \overline{v^2_z},
\label{disprdispz}
\end{equation}

\begin{equation}
\overline{v^2_z} = \frac{1}{\rho} \int^{\infty}_z dz' \rho(R,z')  \frac{\partial \Phi(R,z')}{ \partial z'},
\label{dispzquad}
\end{equation}

\begin{equation}
\overline{v^2_{\phi}} =  \overline{v^2_R} + \frac{R}{\rho} \frac{\partial (\rho \overline{v^2_R})}{ \partial R} + v^2_c,
\label{disphiquad}
\end{equation}

\par\noindent where the circular velocity is $v_c \equiv R \frac{\partial \Phi}{ \partial R}$; with the bars denoting the mean over the quantities in consideration. In the above equations, $\rho$ is the density of the quantity for which we compute the velocity variances, and the potential is due to the whole matter distribution.

From the distribution function $f(R,L_z)$, we conclude that in the azimuthal direction the mean streaming velocity $\overline{v_{\phi}}$ in not necessary null. We, therefore, follow \citet{springel1999}, considering $\overline{v_{\phi}}= f_s v_c$, where generally $f_s (\ll 1)$ is a factor that depends on $\lambda_s$. This means that the streaming velocity of the dark matter halo is a fraction of the local circular velocity. Once specified all the above values, the velocity dispersions for the dark matter halo are given by $\sigma_i = \sqrt{(\overline{v^2_i})}$, with $i=z,R$, and $\sigma^2_{\phi} = \overline{v^2_{\phi}} - \overline{v_{\phi}}^2$.

To the disk, the calculations are similar to the presented above, although the calculation of the $\phi$ component is very different. First, the mean streaming is estimated by the epicyclic approximation, and the equations used to calculate the velocity dispersions are given by

\begin{equation}
\sigma^2_{\phi} = \frac{\sigma^2_R}{\eta^2},
\label{sigmaphidisk}
\end{equation}

\par\noindent where

\begin{equation}
\eta^2 = \frac{4}{R}\frac{\partial \Phi}{\partial R} \left( \frac{3}{R} \frac{\partial \Phi}{ \partial R} + \frac{\partial^2 \Phi}{ \partial R^2} \right)^{-1}.
\label{etaphi}
\end{equation}

Once these quantities are obtained, we use Equation \ref{disphiquad} to evaluate the streaming velocity:

\begin{equation}
\overline{v_{\phi}} = \left(\overline{v_{\phi}^2} - \frac{\sigma^2_R}{\eta^2} \right)^{1/2}.
\label{streamphi}
\end{equation}

In a few words, all this prescription consists in calculating $\sigma_k$,  with $k=R,z,\phi$ for the disk and for the halo at any $(R,z)$ points. To this aim, we employ the following computational techniques.
\begin{itemize}
\item We build a logarithmic mesh, where the density and potentials are calculated at the respective $(R,z)$ points.
\item We use subroutines based on the spline techniques to make the above integrals.
\item We interpolate the dispersions at the particle's points.
\item The particle velocities are set by random numbers from the Schwarzschild's distribution \citep{bt2008}, which is given by
\begin{equation}
f(\vec{v})d^3\vec{v} = \frac{N d^3\vec{v}}{(2 \pi)^{3/2} \sigma_R \sigma_{\phi} \sigma_z} \exp \left[ - \left( \frac{v^2_R}{2 \sigma^2_R} + \frac{v^2_{\phi}}{2 \sigma^2_{\phi}} + \frac{v^2_z}{2 \sigma^2_z} \right) \right],
\label{Schwarzschild}
\end{equation}
where $N$ is the number of particles per unit volume.
\end{itemize}

Following the above prescription, we have written a code to build up a dark matter halo and a baryonic disk particle. We consider both kind of particles gravitationally coupled. We set $N_{\rm{halo}}$ = 30,000 particles for the dark matter halo and $N_{\rm{disk}}$ = 30,000 for the disk in our models. We will see later on that a higher resolution was also used in some  simulations.

The following set of \textit{default} parameters are chosen to realize SdMH disks \citep{sdmh2005}: total mass $M_t = v^2_{200}/(10G\,H_0) = 0.98 \times 10^{12} M_{\odot}$, where $v_{200} = 160 km\,s^{-1}$ is the virial velocity, $G$ is the gravitational Universal constant, $H_0 = 100 km\, s^{-1}\, Mpc^{-1}$ is the Hubble constant; the total mass of the disk $M_{d} = m_d M_t$, where $m_d=0.041$ is a dimensionless fraction of the total mass, the disk scale length $h = 2.74$ kpc, the disk vertical scale-height $z_0 \sim 0.2 h$, and the spin parameter $\lambda = 0.033$.

\begin{figure}
\epsscale{1.0}
\plotone{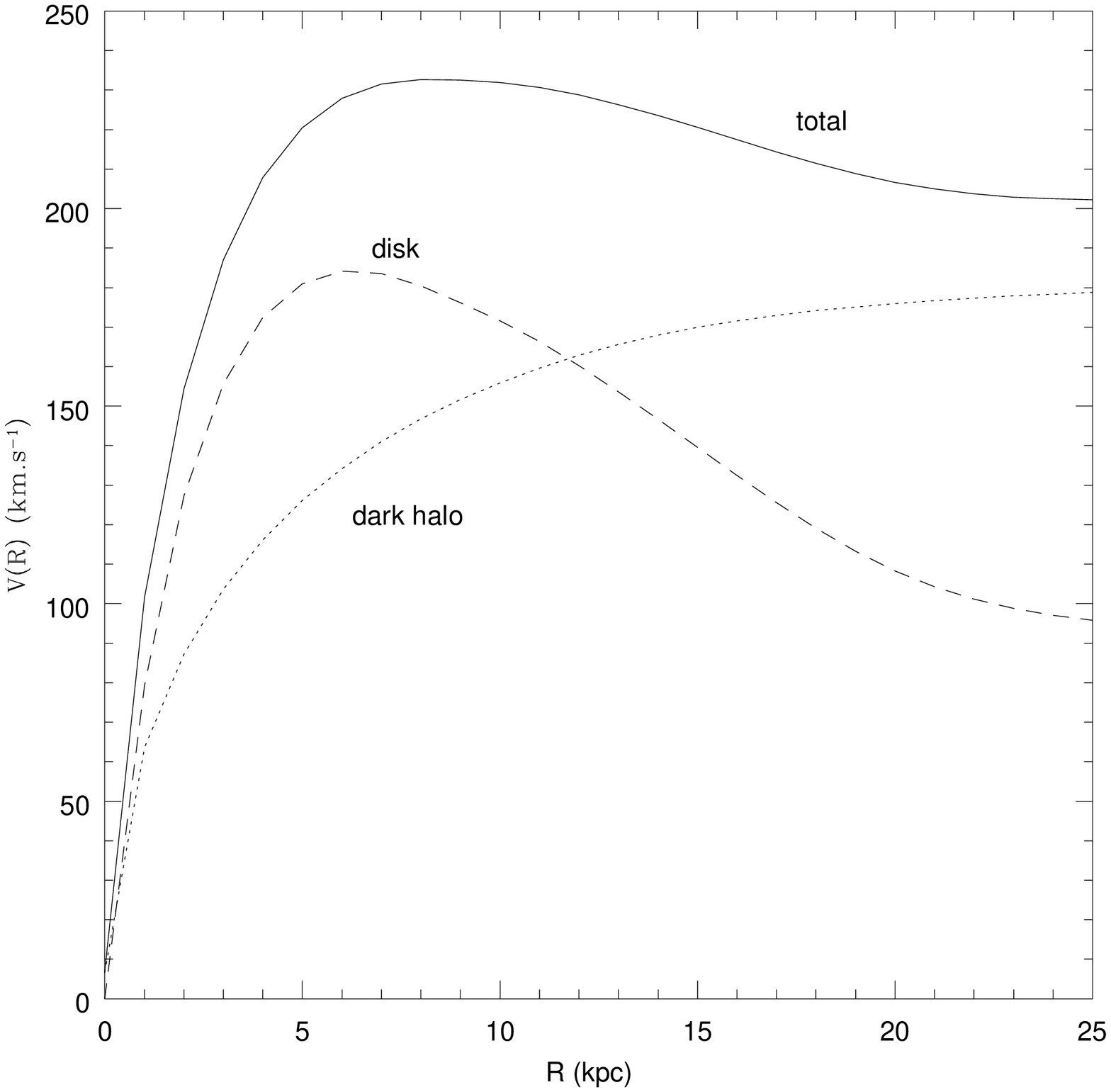} 
\caption{Rotation curves of our galaxy model in cylindrical coordinates. The abscissa shows the distance from the center, in kpc, in the plane of the disk. The ordinate shows the velocities for the disk, the halo, and the whole galaxy.}
\label{rotationcurves}
\end{figure}

In Figure \ref{rotationcurves}, we display the rotation curves of the modeled galaxy, using the parameters described above. We note that our results are very similar to that by \citet{sdmh2005}. The differences reside on the numerical procedures and techniques to compute the disk potentials, particle noise due to the halo, disk truncation, etc.

\section{SIMULATIONS AND DISCUSSION OF THE RESULTS} 
\label{results}

We now present and discuss the results of our simulations. As usual in treecodes, we have chosen the tolerance parameter $\theta=0.8$, which gives a better performance for the calculations. Other typical parameters are the halo smoothing scale length $l_h=0.15$ and the disk smoothing scale length $l_d=0.10$. It is worth noting that we have changed the scale length parameters in the following range: $l_h=0.15 \pm 0.1$ and $l_d=0.10 \pm 0.05$; and our results do not change significantly.

The following simulations are performed: Newtonian potential with the default Gadget-2 code and YGP with $\lambda=$ 1, 10, 100, and 1000 kpc.

The justification for choosing the above values for $\lambda$ is the following. Since spiral galaxies have characteristic dimensions of tens of kpc we choose the Yukawa scale lengths to be much lower (i.e., 1 kpc), much greater (i.e., 1000 kpc), and of the order (i.e., 10 and 100 kpc) of the spiral galaxy sizes, in order to see how the galaxy structure is affected.
It is expected that for $\lambda$ much larger than the characteristic dimensions of galaxies, the YGP model is similar
to the Newtonian one. On the other hand, for $\lambda$ smaller than the characteristic dimensions of galaxies, the YGP and the Newtonian models yield very different results.

\subsection{The Newtonian Simulation} 
\label{newtoniansim}

We use the default Gadget-2 code and make a typical Newtonian simulation with a total simulated time of $t=1$ Gyr. This model is used like a ``control group", with which we compare the other runs.

\begin{figure}
\epsscale{1.0}
\plotone{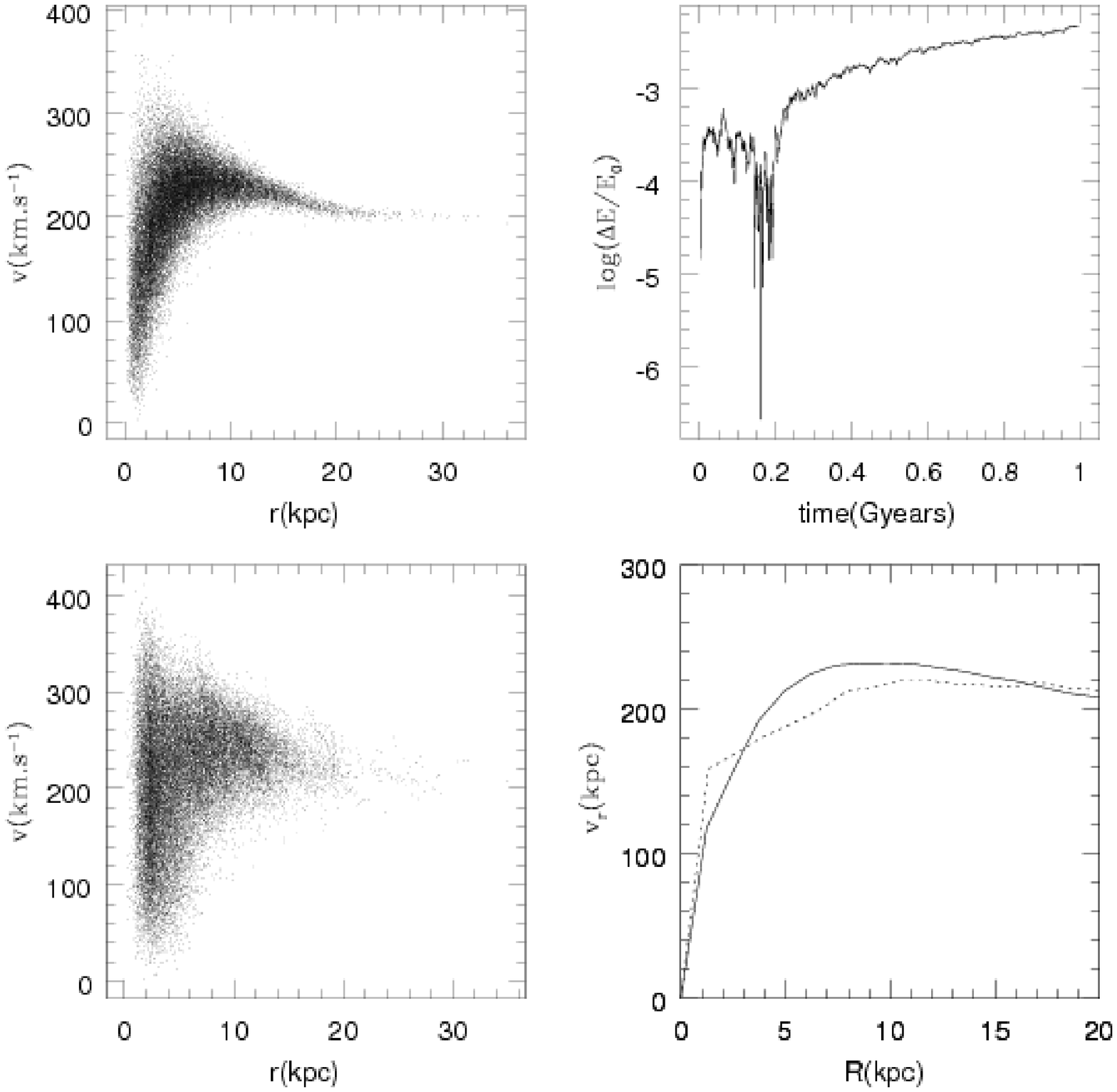} 
\caption{Top left: phase space for initial snapshot data. Top right: energy conservation of the simulation. Bottom left: phase space for final snapshot data at 1 Gyr. Bottom right: Rotation curves for initial (solid line) and final (dashed line) snapshots.}
\label{fignewt1}
\end{figure}

In Figure \ref{fignewt1}, we show the principal features of the Newtonian run. In this figure we display at the top left panel the phase space points $r \times v$ of the initial snapshot, where $r$ is the distance from the center of the matter distribution in kpc and $v$ is the modulus of the velocity, in $\rm{km\,s^{-1}}$;
at top right, we show the relative energy conservation $\Delta E / E_0$, where $E(t)$ is the total energy of the simulated system (disk plus halo) at time $t$,  $\Delta E = E(t) - E(0)$, and $E(0) \equiv E_0$. From this frame, we conclude that the energy conservation violation is less than $1 \%$, showing that our simulation is reliable.
At the bottom left, we present the phase space points at final time, $t = 1$ Gyr; Finally, at bottom right, we present the rotation curve $R \times v_r$, where $R$ is the radial cylindrical coordinate and $v_r$ is the rotation velocity. Note the similarity between the initial and the 1 Gyr rotation curves; one concludes that early-type galaxy structure can be accounted for the Newtonian model quite well, as is well known.

\begin{figure}
\epsscale{1.0}
\plotone{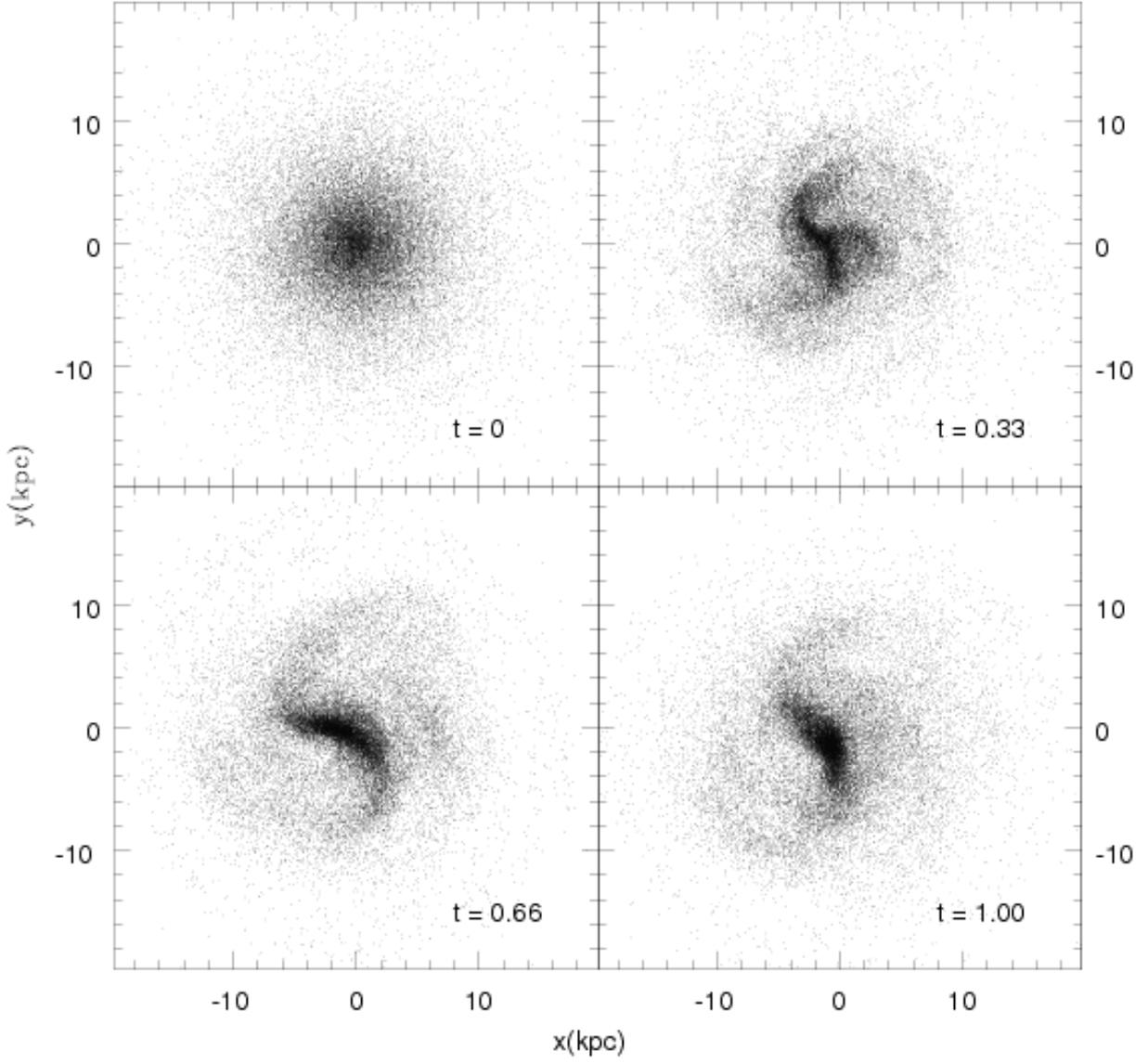} 
\caption{Newtonian disk at $z-$projection at 0, 0.33, 0.66, and 1 Gyr of simulated time (indicated in the respective boxes).}
\label{fignewt2}
\end{figure}

In Figure \ref{fignewt2}, we show the snapshots at $t=0,0.33,0.66$ and 1 Gyr. We note that the disk evolves to a late-type system with a central bar, for which spiral arms are present in each end. Bars usually appear in $N$-body simulations, and their developments in these simulations seem to be connected with the various parameters listed in Section \ref{assembly}. It is worth noting that there is a branch in the Galactic Dynamics which studies the global stability of the differentially rotating disks and the formation of bars \cite[see chapter 6.3 of][]{bt2008}.

It could be argued that some characteristics of our numerical procedures could contribute to the bar formation, but this feature is far from being a bad result. A recent study \citep{verdes}, based on observational data of isolated galaxies, interprets bars and arms as a natural consequence of secular evolution of late-type systems, and this is also corroborated by numerical simulations. These authors conclude that isolated galaxies do not seem to be preferentially barred or unbarred. Therefore, in this work, bars will be considered as a natural feature of the simulated late-type systems.

\begin{figure}
\epsscale{1.0}
\plotone{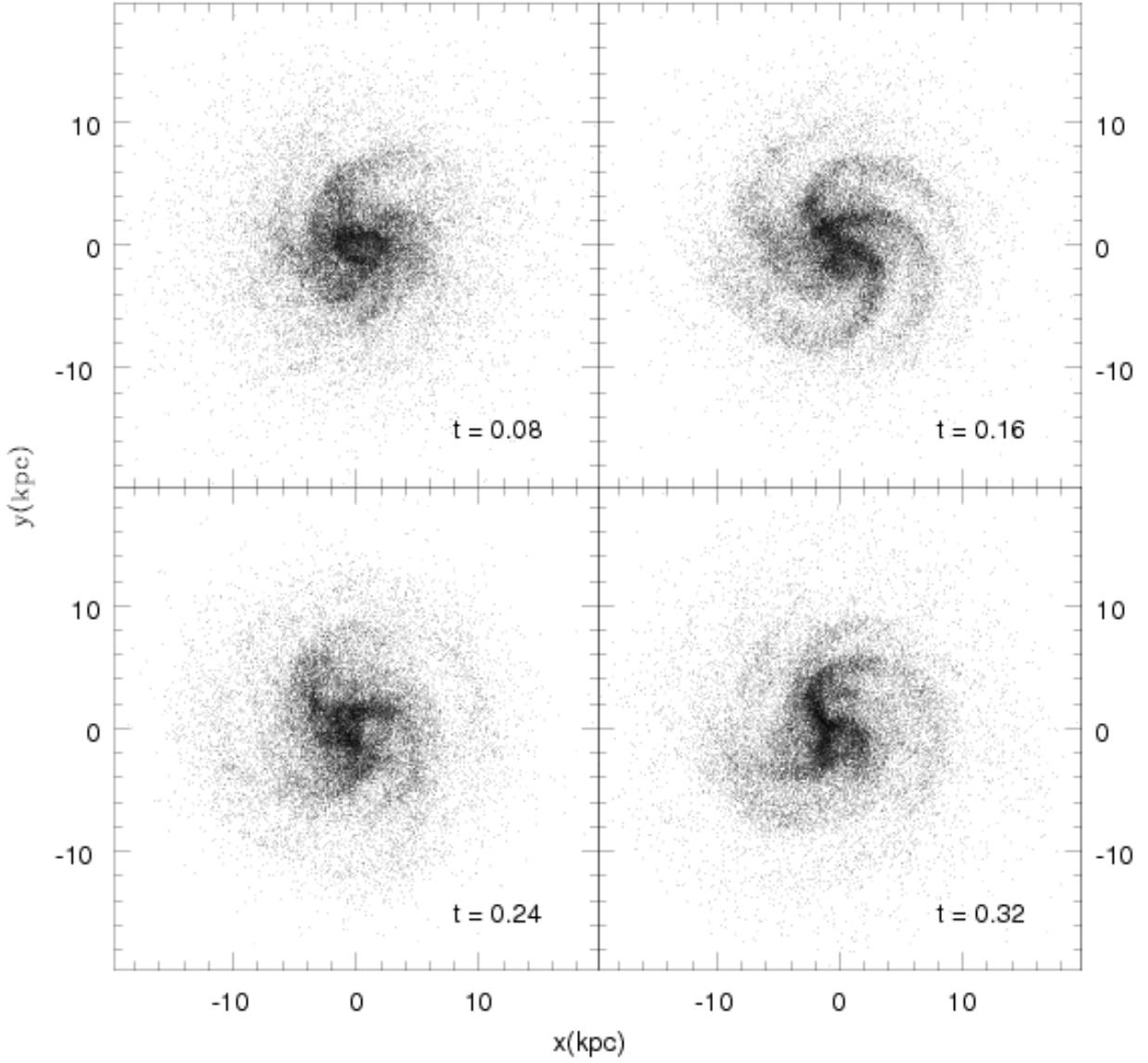} 
\caption{First 320 Myr of simulated time to the Newtonian disk at $z-$projection. Time is indicated in the respective boxes.}
\label{fignewt3}
\end{figure}

In Figure \ref{fignewt3}, we note the development of spiral arms in the first 0.33 Gyr of simulated time, due to swing amplification \citep{bt2008}. This result, as already mentioned, is expected in $N$-body simulations of disk galaxies \citep{sdmh2005}.  We conclude, from all these figures, that our Newtonian simulation yields a typical morphology found in the Cosmos. We emphasize that we built up other models which are stable against bars (mostly the big disk ones), but we chose to consider the Milky-Way like model in the present work.

In the next subsections we simulate YGP late-type galaxies for different values of $\lambda$.

\subsection{The YGP Simulations: $\lambda=100$ and $\lambda=1000$ kpc} 

We run the galaxy model with $\lambda=100$ and $\lambda=1000$ kpc, whose results are remarkably similar. Therefore, we discuss here only the $\lambda=100$ kpc case.

\begin{figure}
\epsscale{1.0}
\plotone{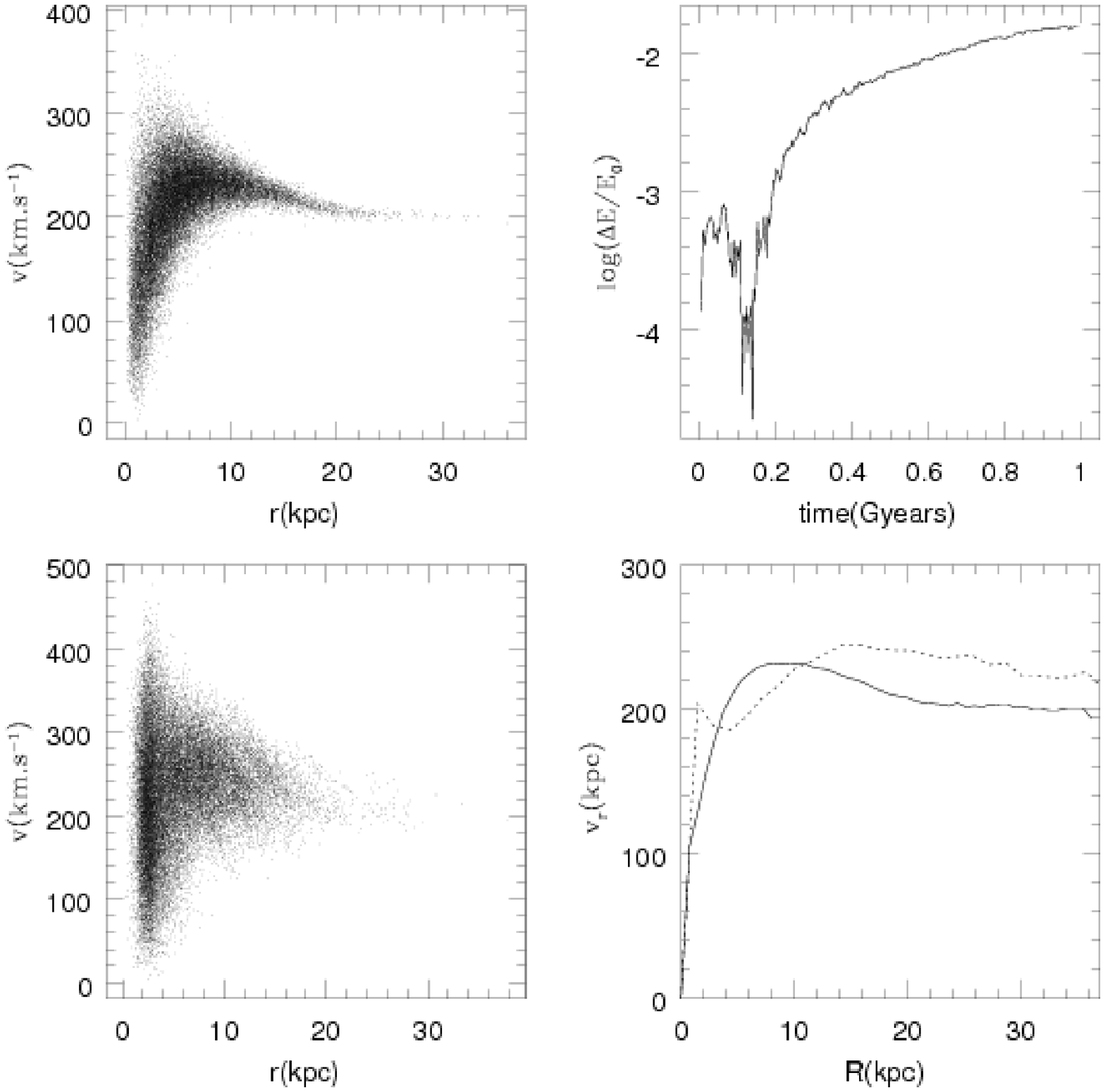}   
\caption{Top left: phase space for the initial snapshot for the Yukawian disk simulation with $\lambda=100$ kpc. Top right: energy conservation of the simulation. Bottom left: phase space for final snapshot data at 1 Gyr.
Bottom right: Rotation curves for the initial (solid line) and the final (dashed line) snapshots. }
\label{yuk100fig1}
\end{figure}

From Figure \ref{yuk100fig1}, we note its resemblance to the Newtonian simulation. In this figure, only the final rotation curve is slightly different from the Newtonian runs, but it is compatible with observational curves. In the bottom left panel of this figure, one can see a peak in the rotation curve around 0 kpc $\le R \le 2$ kpc,  which is due to a bar that rotates like a rigid body.

\begin{figure}
\epsscale{1.0}
\plotone{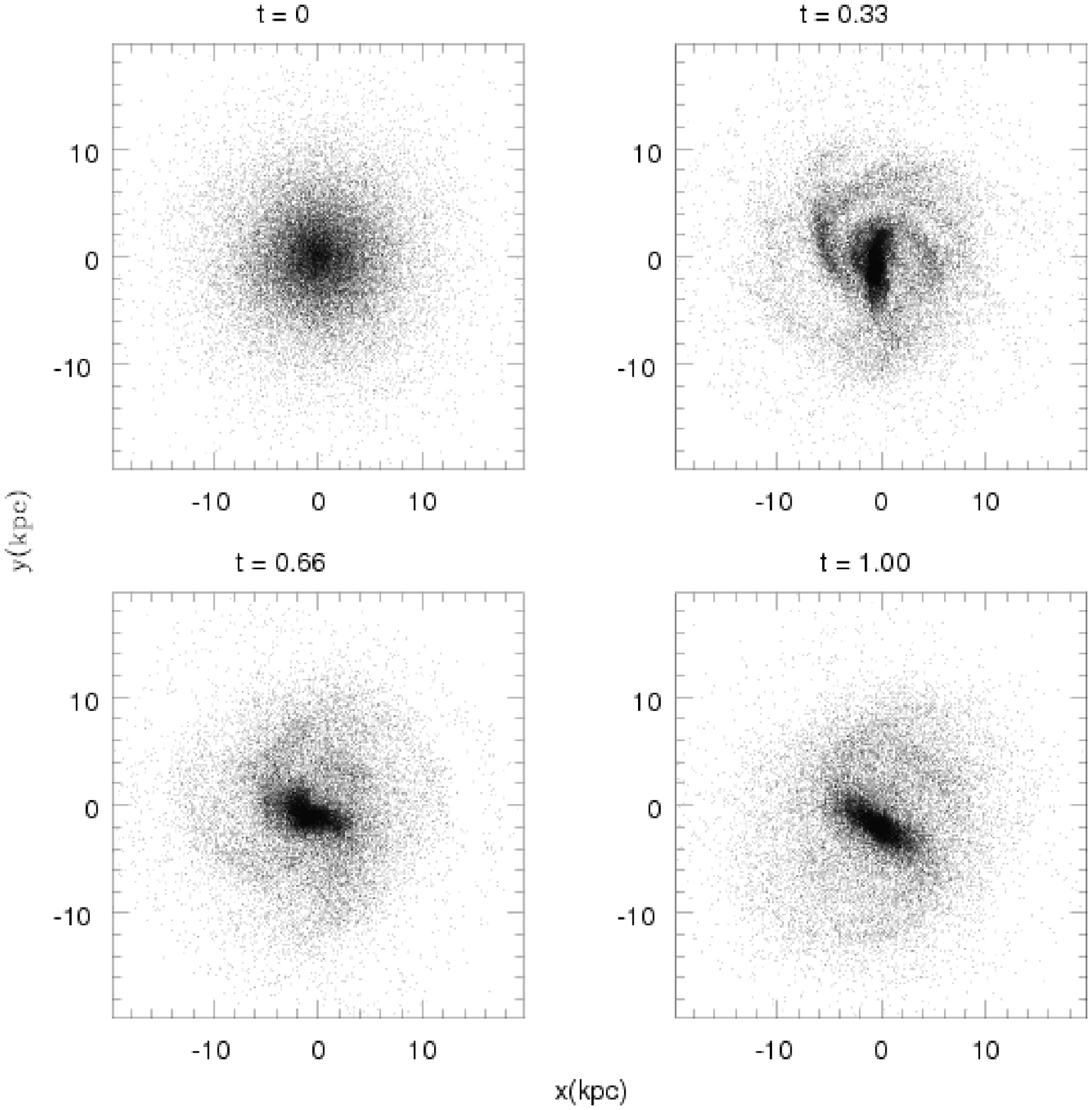}  
\caption{Disk at $z-$projection at 0, 0.33, 0.66, and 1 Gyr of simulated time (indicated in the respective boxes) for $\lambda = 100$ kpc.}
\label{yuk100fig2}
\end{figure}

\begin{figure}
\epsscale{1.0}
\plotone{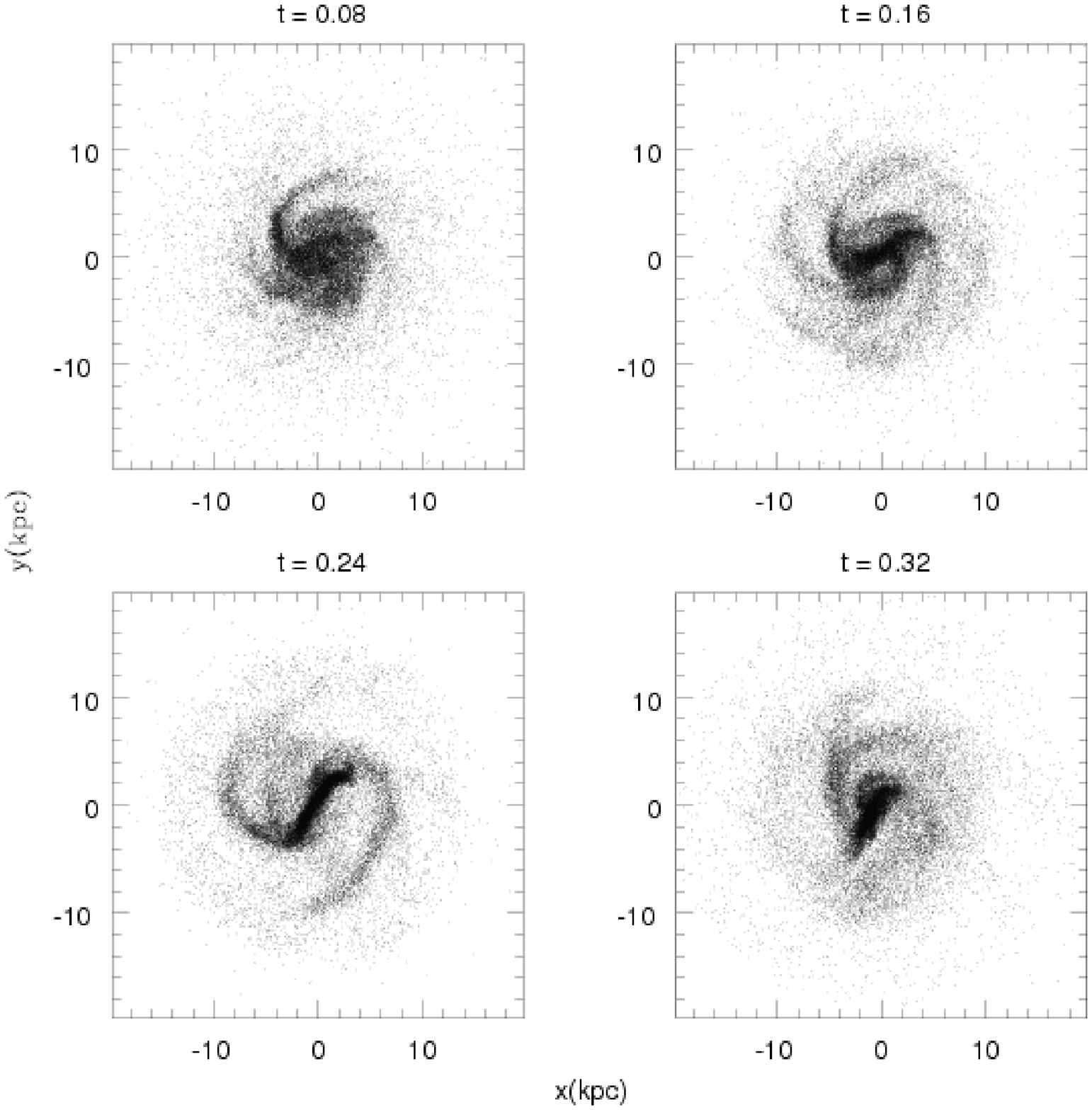} 
\caption{First 320 Myr of simulated time to the Yukawian disk at $z-$projection for $\lambda=100$ kpc. Time is indicated in the respective boxes.}
\label{yuk100fig3}
\end{figure}

The Figures \ref{yuk100fig2} and  \ref{yuk100fig3} display the particle's positions of the disk in the $xy$-plane, at different simulated times. In particular, Figure \ref{yuk100fig2} displays four snapshots for $t=$ 0, 0.33, 0.66 and 1 Gyr. We note that the system evolves to a disk with a central bar and some spiral arms, although these arms are not so remarkable, in comparison to the Newtonian case.

The Figure \ref{yuk100fig3} shows the first 320 Myr of the disk's evolution. We can see at $t=0.08$ and $t=0.16$  Gyr some remarkable spiral arms, which are expected features, due to the swing amplification. We can see that the number of spiral arms changes as the simulated time increases, and the system evolves to a central bar earlier than in the Newtonian case.

Although the $\lambda=100$ kpc simulation describes quite well the morphological aspects of spiral galaxies, it is important to know if some characteristics discussed above, such as the bar for example, depend on the resolution adopted ($\sim 10^4$ particles). We then construct a model with a higher resolution using the same physical parameters, but now with 300,000 particles for the disk and 600,000 for the halo, respectively\footnote{All the higher resolution simulations were performed in the ``HPC Bull Cluster" belonging to the State University of Santa Cruz, which was sponsored by FAPESB.}. The smoothing scale lengths were recalculated and now read $l_h = 0.01$ and $l_d=0.004$, respectively (see Brandao \& de Araujo 2010a, and references therein).

\begin{figure}
\epsscale{1.0}
\plotone{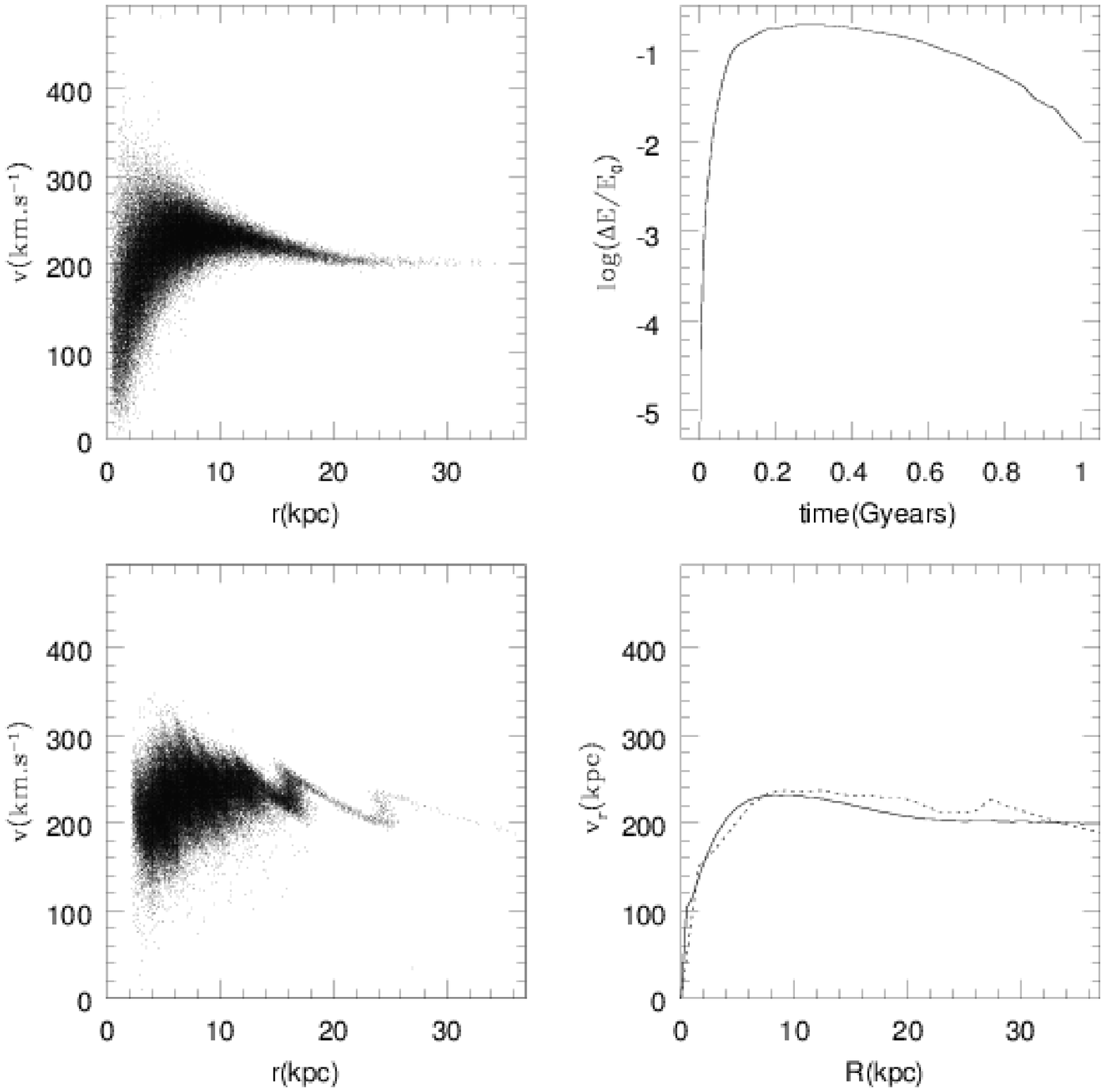}
\caption{Same as in Figure \ref{yuk100fig1} for the high-resolution simulation for $\lambda = 100$ kpc.}
\label{anahigh100}
\end{figure}

\begin{figure}
\epsscale{1.0}
\plotone{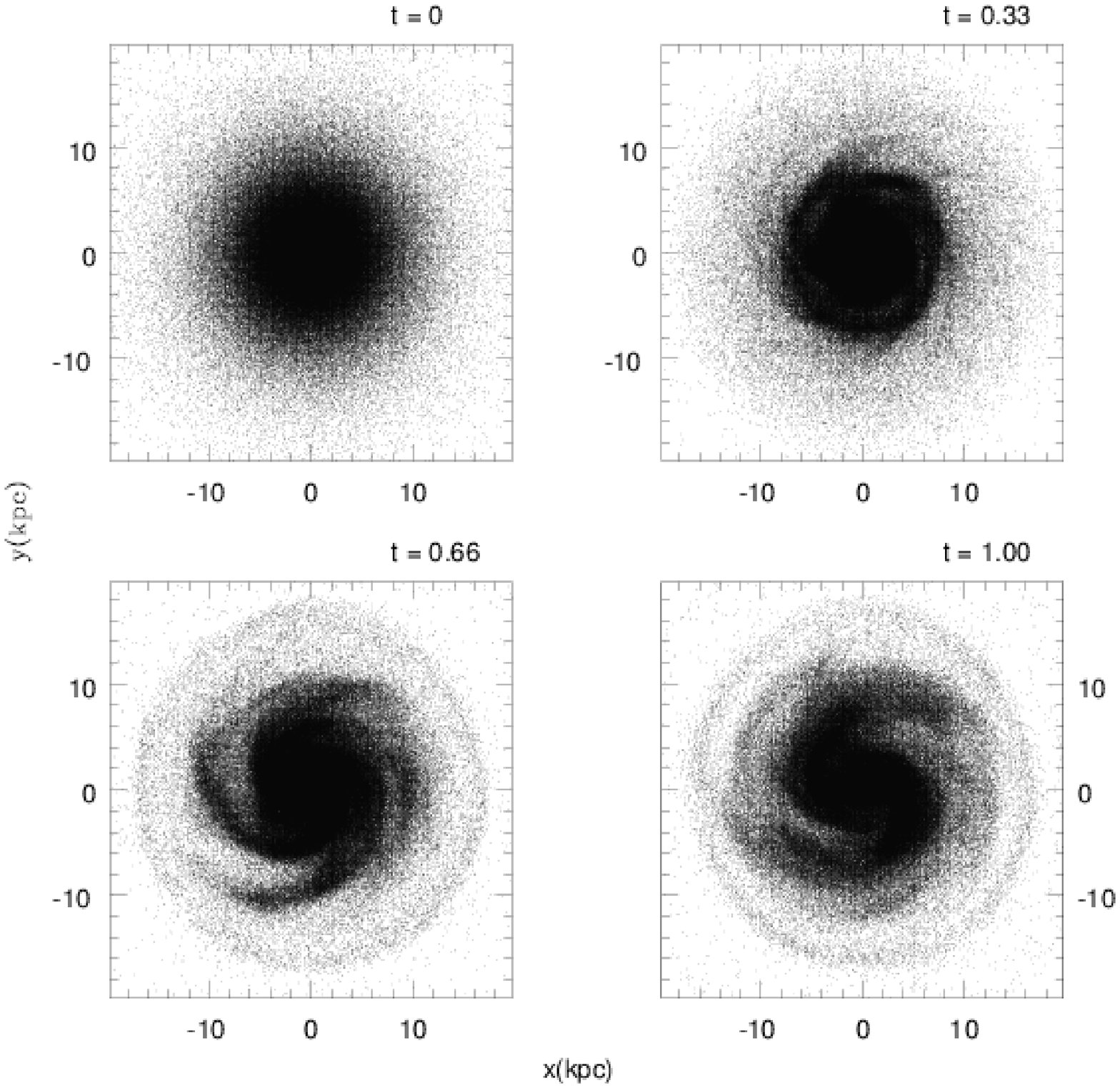} 
\caption{Same as in Figure \ref{yuk100fig2} for the high-resolution simulation for $\lambda = 100$ kpc.}
\label{plothigh100}
\end{figure}

In Figures \ref{anahigh100} and \ref{plothigh100}, we present the results for the high-resolution simulations for $\lambda=100$ kpc. Note that the relative energy conservation at the end of the simulation is $\sim 1 \% $, showing that our simulations are reliable. Comparing these figures with those for lower resolution one notes some similarities, although the higher resolution figures obviously stand out more clearly the features of the modeled galaxy. The bar still appears, showing that it is not an effect related to the low resolution. The rotation curves for both resolutions are also similar, with both resembling a Newtonian rotation curve.

In conclusion, we have seen in this subsection some interesting results that show that the YGP can reliably model a spiral galaxy if $\lambda \gg 10$ kpc. In this case, the YGP presents results that are similar to the Newtonian ones.

\subsection{The YGP Simulation: $\lambda=10$ kpc \label{10kpc}} 

\begin{figure}
\epsscale{1.0}
\plotone{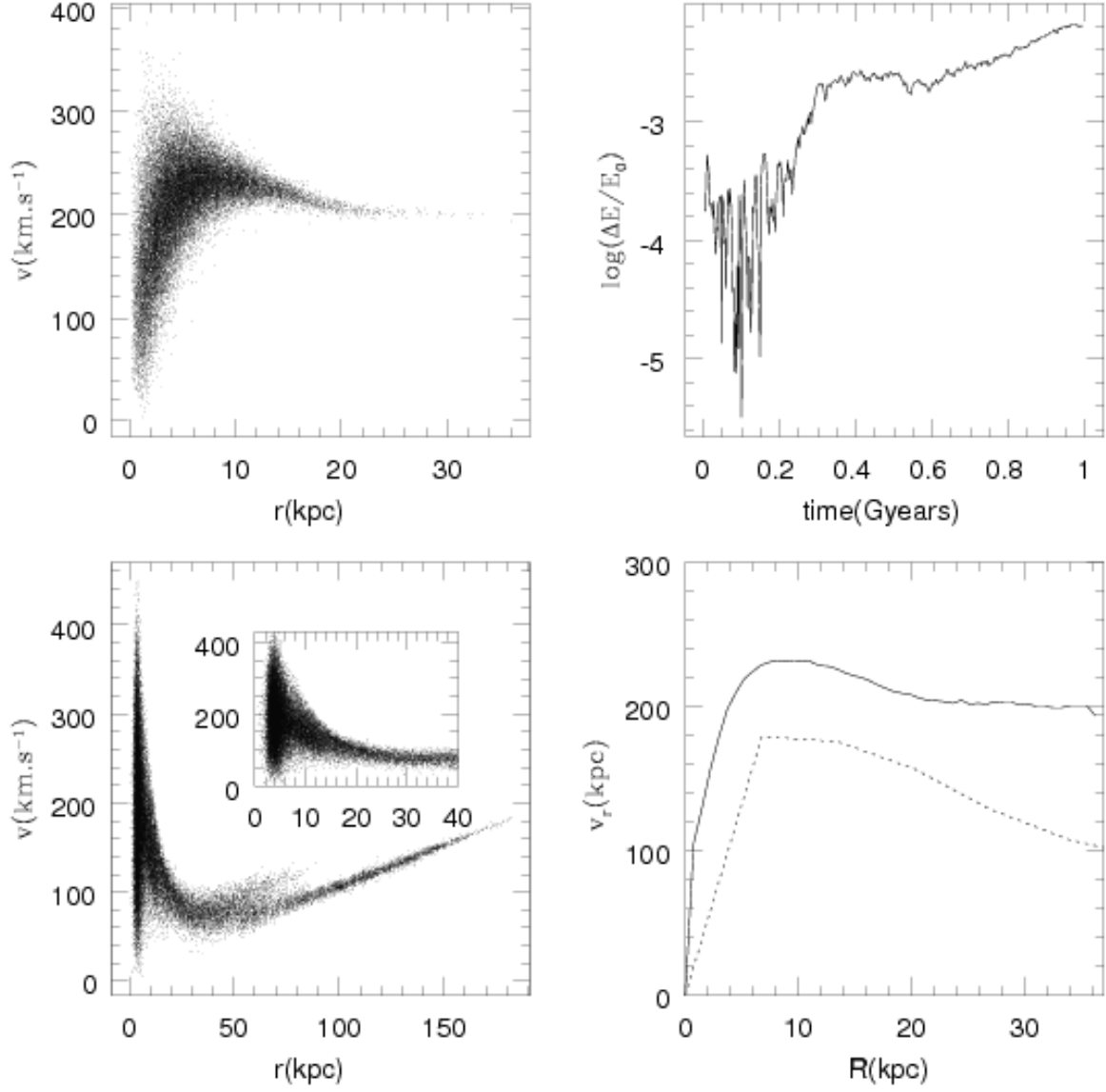}    
\caption{Top left: phase space for the initial snapshot for the Yukawian disk simulation with $\lambda=10$ kpc. Top right: energy conservation of the simulation. Bottom left: phase space for final snapshot data at 1 Gyr. Also shown is a zoom of the final rotation curve, for comparison. Bottom right: Rotation curves for the initial (solid line) and the final (dashed line) snapshots.}
\label{yuk10fig1}
\end{figure}

In Figure \ref{yuk10fig1}, we display the same as in Figure \ref{yuk100fig1} but for $\lambda=10$ kpc. Note that the initial and final rotation curves are quite different. This shows that it is not possible to produce a $\lambda=10$ kpc Yukawian spiral galaxy consistent with the observations.

Note that the energy violation, also shown in Figure \ref{yuk10fig1}, is better than in the case for $\lambda=100$ kpc; this implies, therefore, that our simulations are quite reliable.

\begin{figure}
\epsscale{1.0}
\plotone{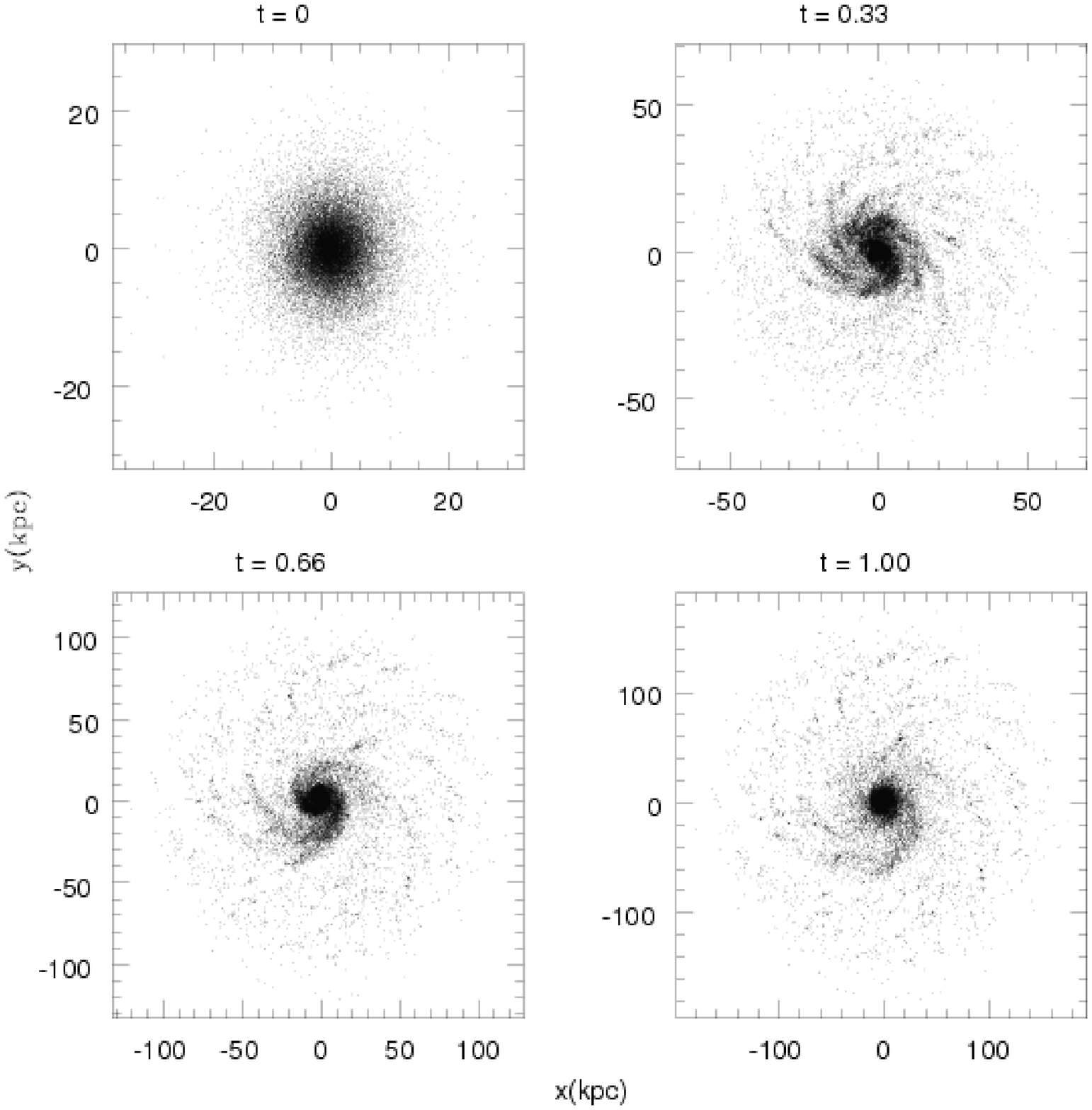}  
\caption{Disk at $z-$projection at 0, 0.33, 0.66, and 1 Gyr of simulated time (indicated in the respective boxes) for $\lambda = 10$ kpc.}
\label{yuk10fig2}
\end{figure}

\begin{figure}
\epsscale{1.0}
\plotone{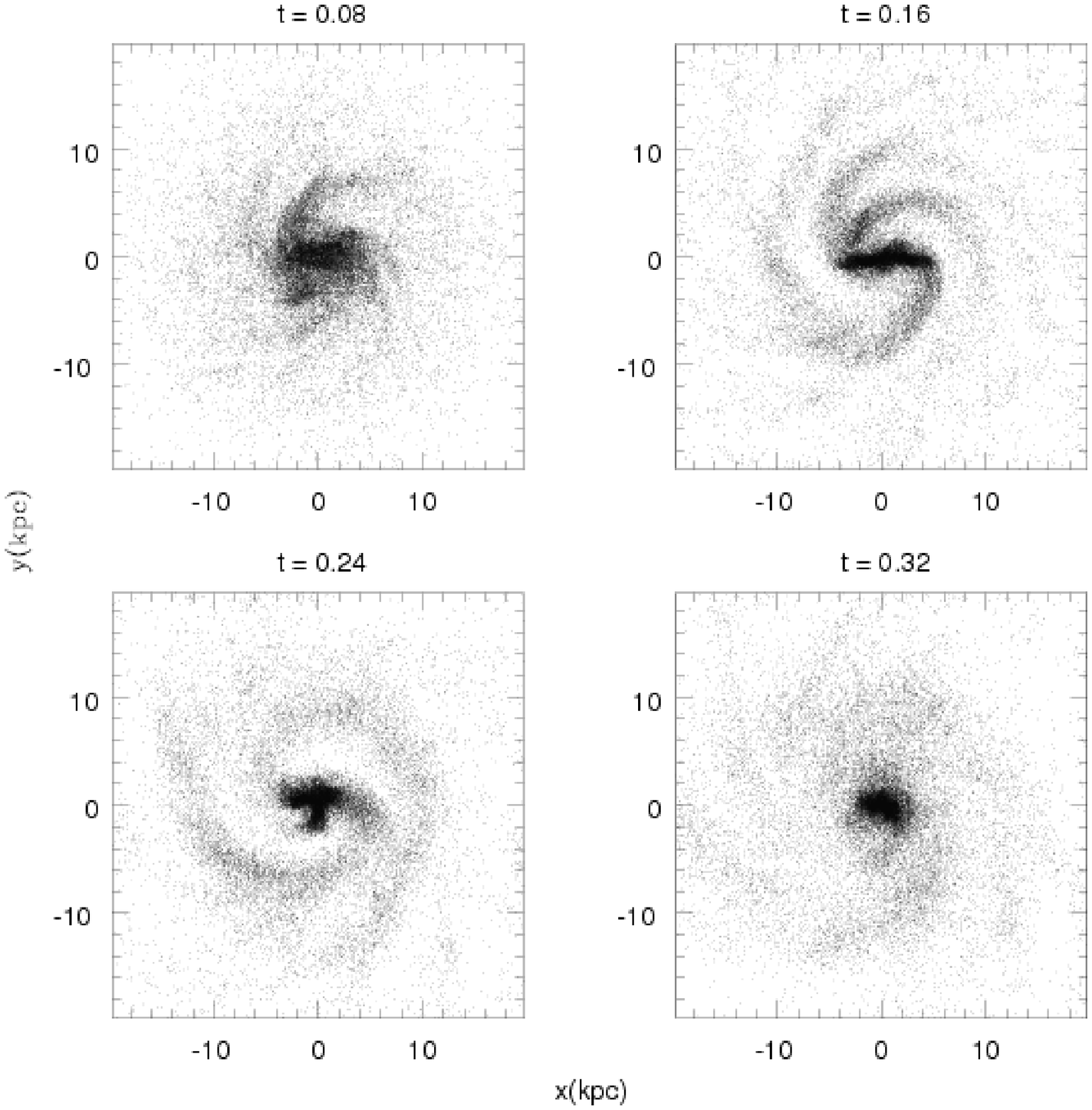} 
\caption{First 320 Myr of simulated time to the Yukawian disk at $z-$projection for $\lambda=10$ kpc. Time is indicated in the respective boxes.}
\label{yuk10fig3}
\end{figure}

In Figure \ref{yuk10fig2}, we show how would be a $\lambda=10$ kpc Yukawian spiral galaxy.  In this sequence of snapshots, we can see the growth of the disk as well as the evolution of the central part of the galaxy. It is interesting to note that the core becomes smaller than initially. The YGP for $\lambda=10$ kpc makes the central parts of the disk shrink. The first 320 Myr of the simulation is shown in Figure \ref{yuk10fig3}. We see again the swing amplification and the development of spiral arms.

\begin{figure}
\epsscale{1.0}
\plotone{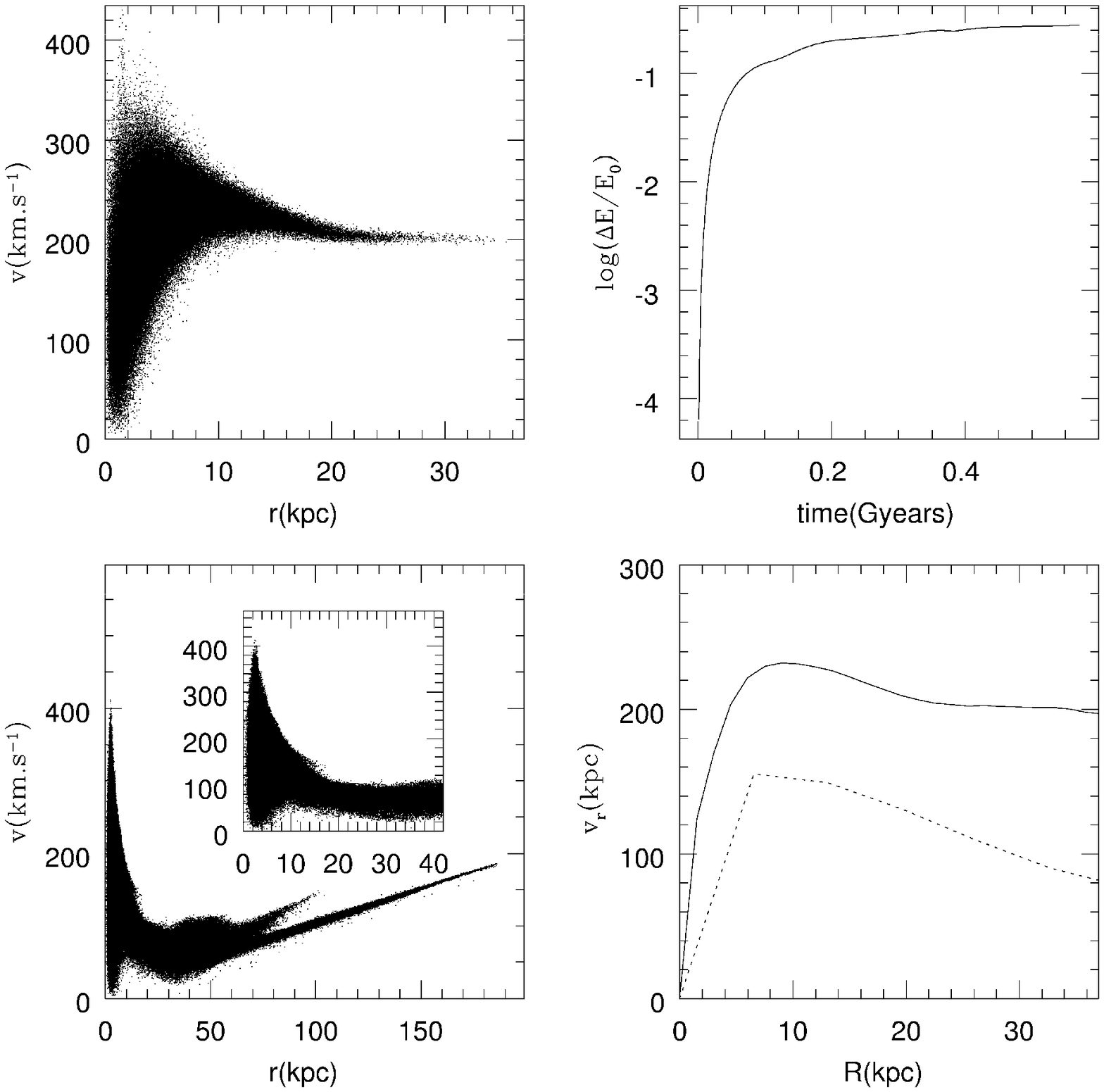} 
\caption{Same as in Figure \ref{yuk10fig1} for the high-resolution simulation for $\lambda = 10$ kpc.}
\label{anahigh10}
\end{figure}

\begin{figure}
\epsscale{1.0}
\plotone{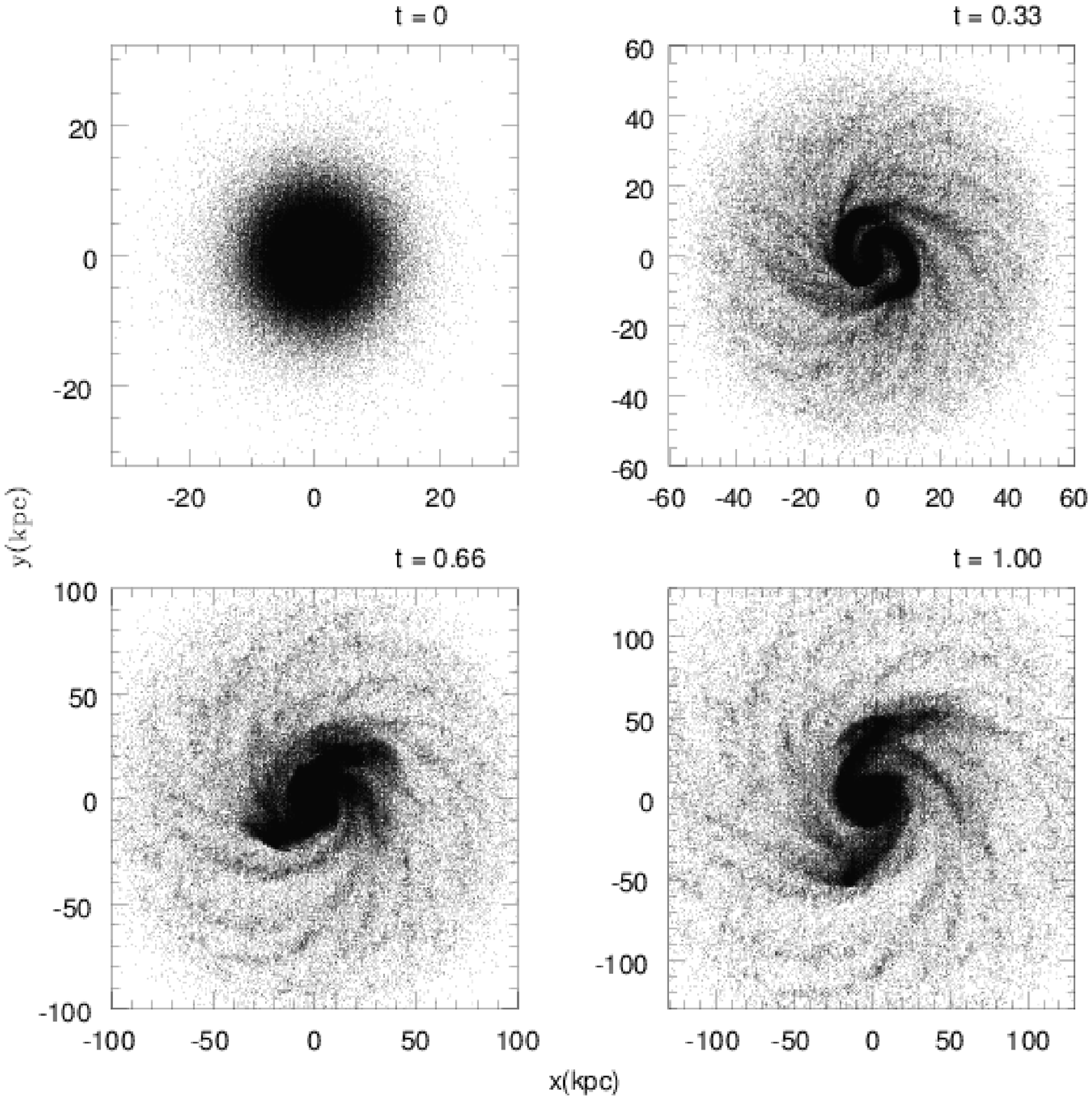} 
\caption{Same as in Figure \ref{yuk10fig2} for the high-resolution simulation for $\lambda = 10$ kpc.}
\label{plothigh10}
\end{figure}

Similarly to what we did for the simulations described in the previous subsection, we also simulate a high-resolution model for $\lambda=10$ kpc. Figures \ref{anahigh10} and \ref{plothigh10} show the results of the high-resolution simulations. As in the case for $\lambda=100$ kpc, the relative energy conservation at the end of the simulation is $\sim 1 \% $ for $\lambda=10$ kpc, showing that our simulations are reliable. As before, comparing these figures with those for lower resolution one notes that the high resolutions  obviously stand out more clearly the features of the modeled galaxy, but the main conclusions presented above are the same. In particular, the rotation curves for both resolutions are similar, but they are quite different from a Newtonian rotation curve.

Summing up a YGP with $\lambda=10$ kpc does not model a spiral galaxy appropriately.

\subsection{The YGP Simulation: $\lambda=1$ kpc} 

\begin{figure}
\epsscale{1.0}
\plotone{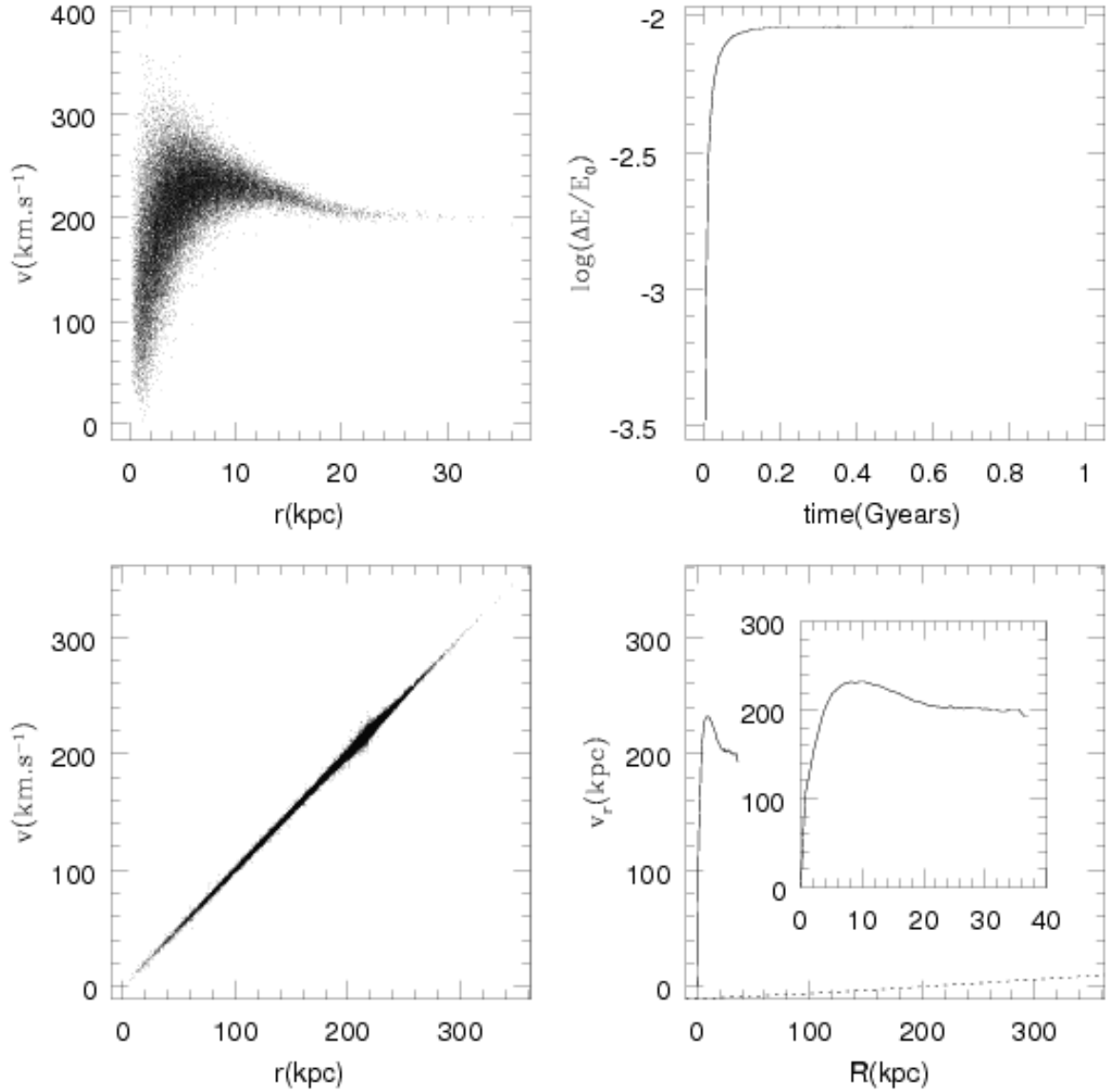}   
\caption{Top left: phase space for the initial snapshot for the Yukawian disk simulation with $\lambda=1$ kpc. Top right: energy conservation of the simulation. Bottom left: phase space for final snapshot data at 1 Gyr. Bottom right: rotation curves for the initial (solid line) and the final (dashed line) snapshots. Also shown is a zoom of the initial rotation curve for comparison.}
\label{yuk1fig1}
\end{figure}

\begin{figure}
\epsscale{1.0}
\plotone{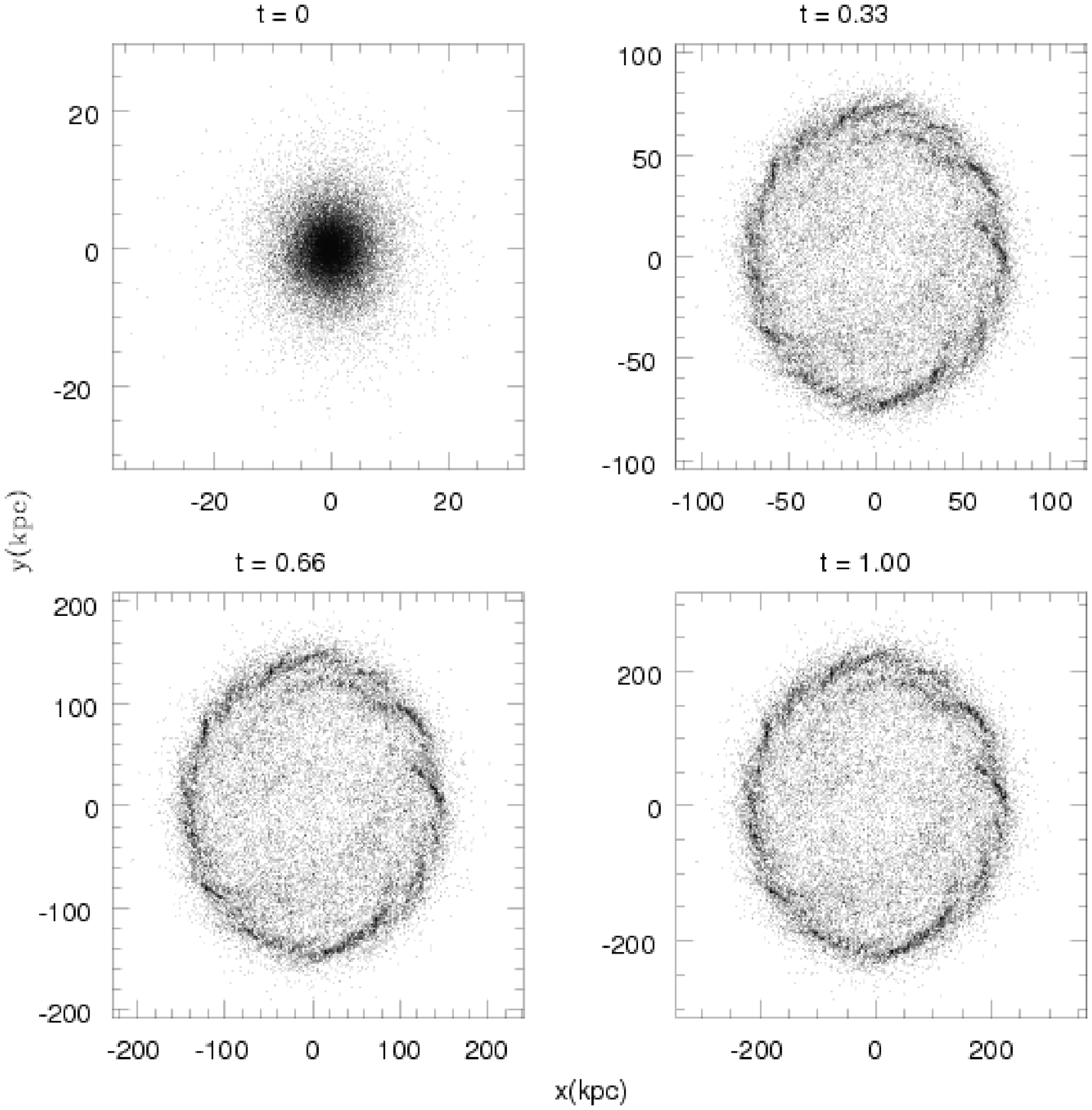} 
\caption{Disk at $z-$projection at 0, 0.33, 0.66, and 1 Gyr of simulated time
(indicated in the respective boxes) for $\lambda = 1$ kpc.}
\label{yuk1fig2}
\end{figure}

\begin{figure}
\epsscale{1.0}
\plotone{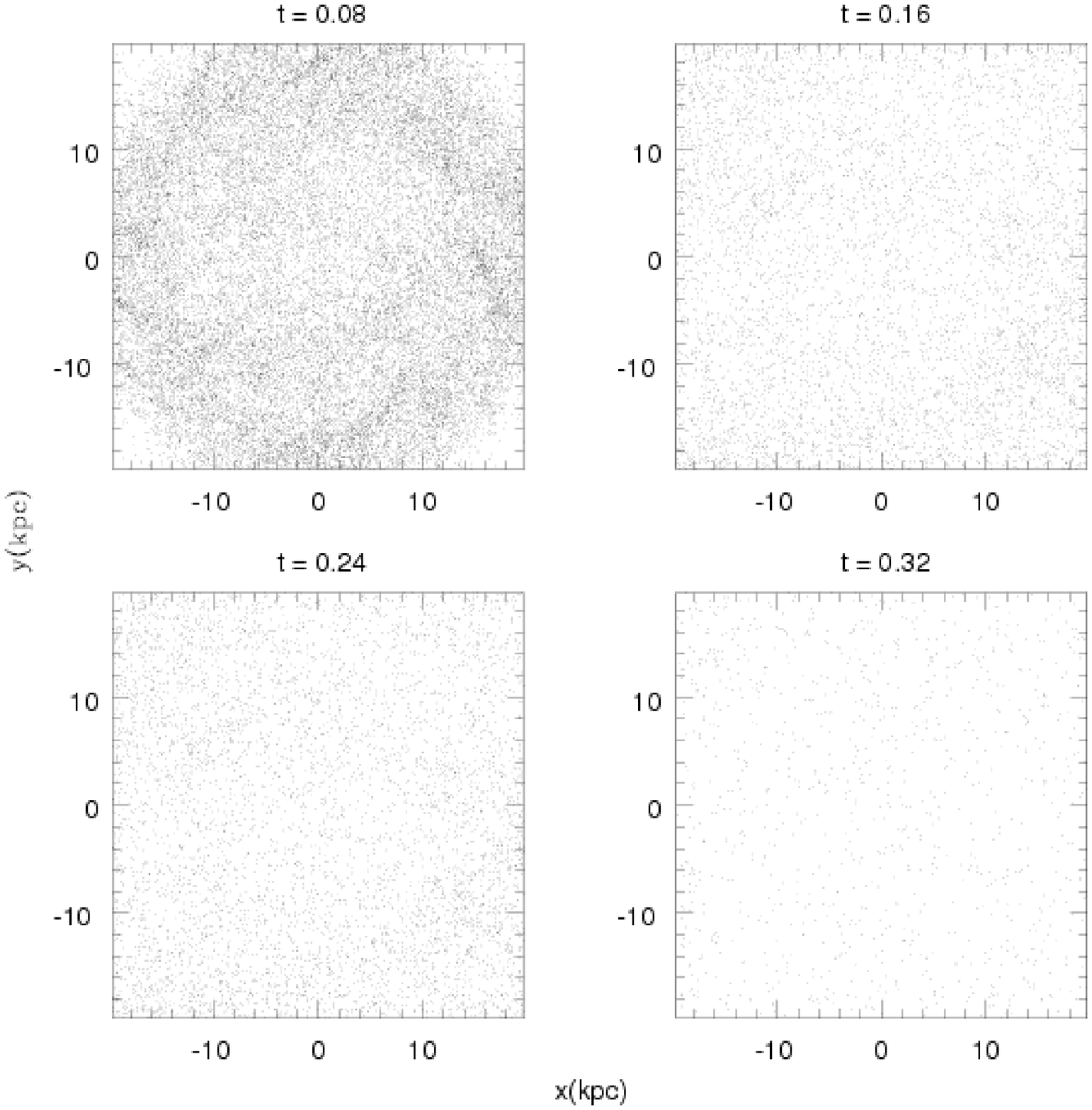} 
\caption{First 320 Myr of simulated time to the Yukawian disk at $z-$projection for $\lambda = 1$ kpc. Time is indicated in the respective boxes.}
\label{yuk1fig3}
\end{figure}

When a spiral galaxy is submitted to the YGP, atypical morphologies appear, in particular for small values of $\lambda$. Figures \ref{yuk1fig1}-\ref{yuk1fig3} display  the results for the Yukawian disk galaxy simulation for $\lambda=1$ kpc.

In Figure \ref{yuk1fig1}, the top right panel shows us that energy violation is very small, proving the reliability of this simulation. The top left panel shows that the initial positions and velocities of the particles in the phase space resemble those of the  Newtonian simulations because we are using the same initial snapshot. But, when the YGP and its corresponding acceleration are considered, their exponential factor plays a decisive role: for $\lambda=1$ kpc, the forces between distant particles ($\gg$ 1 kpc ) are almost ``turned off''. These particles become almost free and, as a result, their initial velocities are almost unchanged. As time goes on, the average distance between particles increases, these particles leave the galaxy and the system becomes more and more diffuse. Faster particles reach distant regions first, and the slower ones spend more time to move through the galaxy. The phase space of the final snapshot is displayed at the bottom left panel and shows us that the particles escape from the galaxy and the initial information is lost. The bottom right panel shows us that the initial rotation curve is lost, namely, the initial and final rotation curves are quite different. From these considerations, we conclude that a putative $\lambda=1$ kpc Yukawian spiral galaxy is ruled out, since it does not resemble the spiral galaxies observed in the universe.

In Figure \ref{yuk1fig2} we show a face-on view of the simulated galaxy, which is gradually destroyed and dispersed in the intergalactic environment. See also Figure \ref{yuk1fig3}, which shows what is happening with the central part of the disk. As time goes on, the central part becomes more and more empty of particles. This result leads one to conclude again that a $\lambda=1$ kpc Yukawian spiral galaxy is ruled out. Although not shown, a high-resolution simulation corroborates these conclusions.

It is worth recalling that \cite{araujo2007} showed that depending on the ratio between $\lambda$ and the scale length of the disk, the disk is destroyed. Such a result is in full agreement with our simulations for $\lambda=1$ kpc.

\subsection{The Surface Density Profile of the Disks of the Simulated Spiral Galaxies} 

So far we have investigated the resulting morphology of our galaxy models via $N$-body simulations. However, this procedure is somewhat incomplete because it is necessary to recover some observational counterparts from the simulated systems in order to test the reliability of our models.

One of these counterparts is just the disk density profiles. In spite of numerical noise, very common in $N$-body simulations, one expects that, in the cases where secular equilibrium plays an important role, the simulated systems maintain their density profiles, even when the mixing in the phase space occurs.

\begin{figure}
\epsscale{1.0}
\plotone{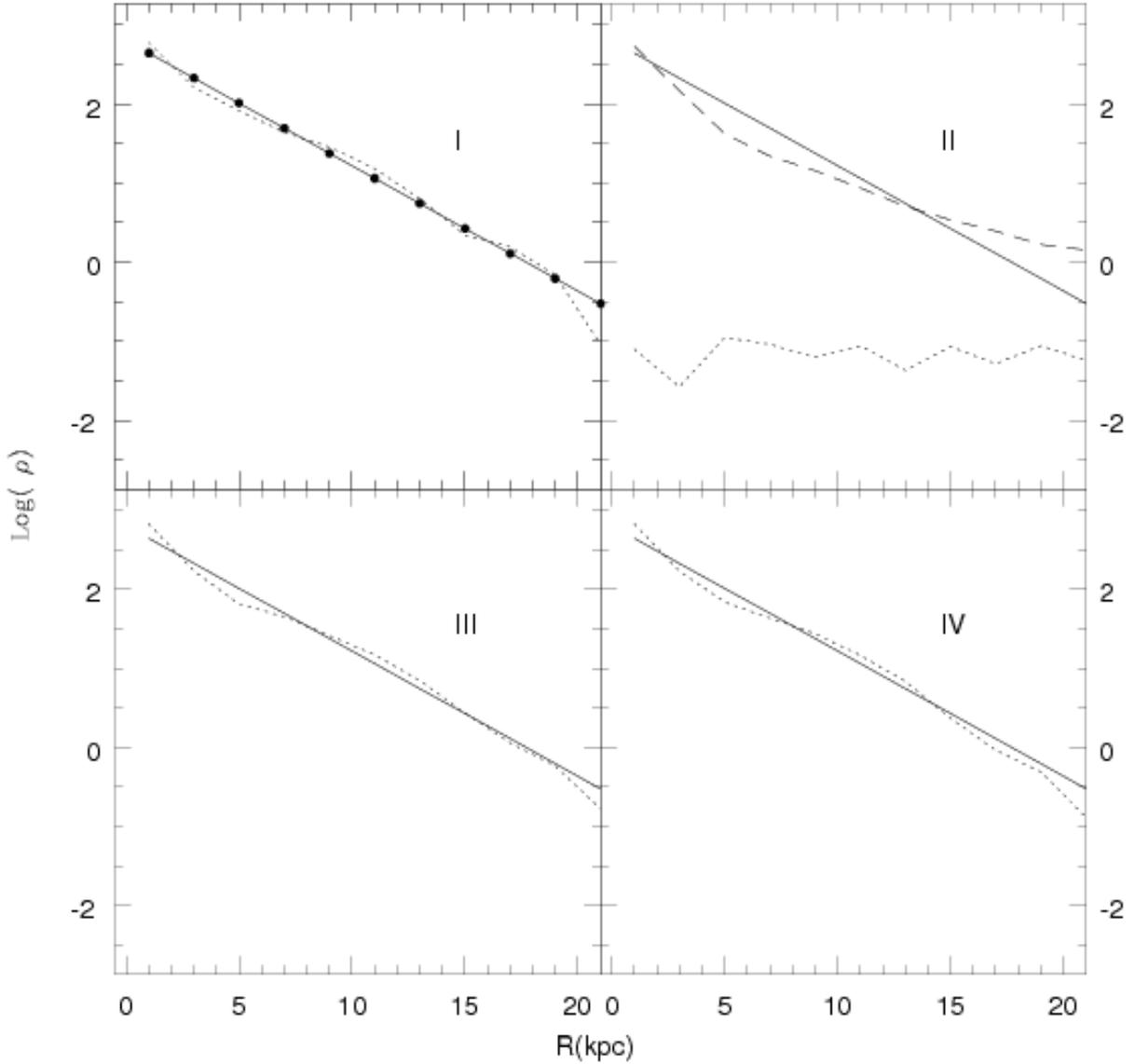} 
\caption{(I) Analytical exponential profile (solid line). Points represent the particle counts per unit area, obtained from the initial snapshot. The dotted line represents the radial profile of the final (at 1 Gyr) snapshot for the Newtonian run. (II) The initial profile (solid line), the final profile for $\lambda=1$ kpc run (dotted line), and the final profile for the $\lambda=10$ kpc (dashed line). (III) The initial profile (solid line) and the final profile for $\lambda=100$ kpc (dotted line). (IV) The initial profile (solid line) and the final profile for $\lambda=1000$ kpc (dotted line).}
\label{yukdensity}
\end{figure}

Then, we calculate from the final snapshots ($t$=1 Gyr), the density profiles of the disks in all cases presented in Sections 3.2-3.4. In Figure \ref{yukdensity}, we display the final density profiles of the simulated disks and compare them with the initial profile. The method used here considers that the disk is composed by concentric rings in the $xy$-plane, where we count the $N^{\star}$ particles in each annulus and apply the formula $\rho = N^{\star} / A$,  where $A$ is the ring's area.

Figure \ref{yukdensity} shows that our Newtonian model is stable, even for 1 Gyr of simulated time, since the initial and final density profiles are almost the same and present an exponential shape.

On the other hand, spiral galaxies modeled with the YGP preserve the exponential disk profile only for $\lambda \gtrsim 100$ kpc. This lower limit for $\lambda$ was also obtained in our previous work with elliptical systems \citep{brandaoearaujo2009b}.

For $\lambda < 100$ kpc there is not an exponential density profile at the end of the simulation.  For example, for $\lambda=1$ kpc the exponential disk is completely destroyed. This result is consistent with the semi-analytical approach by \citet{araujo2007}.

It is worth noting that the density profiles for the high- and low-resolution simulations are similar, that is why we show in Figure \ref{yukdensity} only the results for the low resolution.

From our simulations, it is possible to conclude that the graviton mass is $m_g \ll 10^{-60}$ g, in agreement with \citet{brandaoearaujo2009b}.

The density profile diagnosis then helps one to show how to constrain alternative gravitational potentials, since not necessarily the simulated images of a spiral galaxy and nor its rotation curve allow to constrain.

\section{CONCLUSIONS} 

We give a recipe to probe alternative theories of gravitation in the \textit{non-relativistic r\'egime}, using $N$-body simulations to model spiral galaxies. As an example, we use the recipe given here for the YGP. It is worth stressing
that this recipe can be applied to any other alternative theory of gravitation in which the superposition principle is valid.
In fact, in forthcoming studies we will probe other alternative potentials using the recipe given here.

Basically, one has just to modify the code where the gravitational potential and the corresponding gravitational acceleration are taken into account. Then, one has to model a galaxy, which is initially made consistent with observations and run the simulation with the modified (non-Newtonian) code. For the alternative theory be reliable, the simulated galaxy at, say, 1 Gyr should resemble a true galaxy, i.e., it must have, for example, a disk consistent with that inferred from observations. Models presenting the destruction of galaxies or presenting bizarre morphologies would mean that the potential under consideration would be unreliable.

Although \citet{araujo2007} studied, for example, disk galaxies under YGP, they used analytical arguments (centrifugal equilibrium), while our galaxies behave like ``living" systems because they are composed by thousands of self-gravitating particles. In this way, this can be considered a reliable and strong test due to the fact that $N$-Body systems are very sensitive to chaos and complex phenomena \citep{bt2008}.

In this work, we have studied some models of late-type systems to probe the YGP and to constrain the Yukawian $\lambda$ parameter. As expected, we have seen that if $\lambda$ is much larger than the characteristic dimensions of spiral galaxies, the YGP and the Newtonian models yield the same results. On the other hand, for $\lambda$ smaller than the characteristic dimensions of galaxies, the YGP and the Newtonian models yield very different results. Moreover and more importantly, YGP galaxies for small values of $\lambda$ do not reproduce the rotations curves and the surface density profiles observed in spiral galaxies.

As a general conclusion, if YGP were reliable we should have $\lambda \gtrsim 100$ kpc, otherwise, we could not see late-type systems in the universe. This value of $\lambda$ is larger than that inferred from the solar system's constraints and it could be considered a good estimative, since with such a value the simulated galaxies remain ``alive" for billions of years and look like their observational counterparts.

Last but not the least, we intend in the near future to follow an interesting suggestion given by an anonymous referee, namely, instead of starting our simulations with a Newtonian model we will start with collapsing halos under the Yukawa potential to investigate the characteristics of the equilibrium halo profile, in particular if it resembles to some of the halo profiles discussed in the literature.

\acknowledgments 
C.S.S.B. and J.C.N.A. thank the Brazilian agencies CAPES and FAPESP for support. J.C.N.A. also thanks the Brazilian agency CNPq for partial support. C.S.S.B. also thanks Nick Gnedin for providing the visualization software IFRIT and Volker Springel for the exchange of e-mail messages concerning the realization of disks. Finally, we thank the referee for the careful reading of the paper, the criticisms, and the very useful suggestions which greatly improved our paper.

\clearpage


\clearpage

\clearpage

\clearpage

\clearpage

\clearpage

\clearpage

\clearpage

\clearpage

\clearpage

\clearpage

\clearpage

\clearpage

\clearpage

\clearpage

\clearpage

\clearpage

\clearpage

\end{document}